\documentclass[useAMS,usenatbib]{mn2e}
\usepackage{graphicx}

\def\xmm{{\it XMM-Newton}}
\def\sax{{\it BeppoSAX}}
\def\xte{{\it RXTE}}

\def\ergpscmps{erg\,cm$^{-2}$\,s$^{-1}$}

\title[The Iron line in XTE J1650-500]
      {Iron-line and continuum flux variations in the \xte~ spectra of the black-hole candidate XTE J1650--500}

\author[Rossi et al.]
       {Sabrina Rossi$^{1,2}$\thanks{E-mail: srossi@merate.mi.astro.it},
        Jeroen Homan$^{3}$, Jon M. Miller$^{4}$, Tomaso Belloni$^{1}$ \\
$^{1}$Osservatorio Astronomico di Brera, Via E.~Bianchi 46, 23807, Merate (LC), Italy\\
$^{2}$Universit\` a degli Studi di Milano-Bicocca\\
$^{3}$Center for Space Research, Massachusetts Institute of Technology, 70 Vassar Street, Cambridge, MA 02139, USA\\
$^{4}$Harvard-Smithsonian Center for Astrophysics, 60 Garden Street, Cambridge, MA 02138, USA\\
}
\pagerange{\pageref{firstpage}--\pageref{lastpage}}
\begin{document}
\maketitle
\label{firstpage}

\begin{abstract} We present the results of spectral fits made to 57
pointed observations of the Galactic black hole candidate and X-ray
transient XTE J1650$-$500, made with the {\it Rossi X-ray Timing Explorer}
in 2001 when the source was in a transition from the hard state to the soft state.  A strong and variable Fe~K$\alpha$
emission line is detected in these spectra. The line flux varies in
a non-linear way with the hard X-ray flux, in apparent contradiction to the predictions of simple
disc reflection models. We observe a change in the overall trend that coincides with changes in the continuum X-ray spectrum and the fast X-ray
variability.  The Fe-line flux
versus hard X-ray flux variations are consistent with the predictions of reflection models which consider high disc-ionization states and with a model which considers gravitational light-bending effects. Indications for an anti-correlation between the Fe-line flux and the hard X-ray flux in the spectrally hardest observations and weak variations in the Fe-line energy (as observed with \xmm\ and \sax) slightly favour the light-bending interpretation. \end{abstract}


\section{INTRODUCTION}

The energy spectra of several Galactic black-hole candidates and active
galactic nuclei (AGN) show evidence of broad, relativistic Fe
emission lines \citep{tanafa1995,mifawi2002a,mifain2002,mifawi2002b,mamaka2002,favana2002}. In the framework of
reflection models, these Fe lines are produced as the result of  the
reprocessing of hard X-ray emission by cold ($T<10^6$ K) material in an accretion disc \citep{gefa1991,reno2003}. In the simplest versions of the reflection models, variations in the line strength are expected to correlate
linearly with those in the hard X-ray flux. However, the spectral
variability of some AGN shows that the Fe line does not always
directly trace variations in the irradiating hard X-ray flux. The best known examples of such behaviour have come from {\it ASCA}, \xte, and \xmm\ observations of the Seyfert 1 galaxy MGC-6-30-15 \citep{lefare2000,shiwfa2002,favana2002}.
 
Two explanations have been proposed for the behaviour observed in 
MGC-6-30-15 with {\it XMM-Newton}. The first one \citep{bavafa2003} assumes reflection of hard X-ray emission by a strongly photoionized accretion disc and relies strongly on the ionized disc models of \citet{rofa1993}. Simulations have shown that the shape of the reflected spectrum is
sensitive to the ionization state of the disc \citep{rofa1993,rofayo1999}, its density profile
\citep{nakaka2000,barofa2001} and the properties of the illuminating radiation \citep{baro2002}. \citet{baro2002} showed that the energy,
the equivalent width and the flux of the Fe line are expected to
depend strongly on the ratio of illuminating flux to disc flux and
the photon index of the irradiating emission (which is modelled as a power-law). In particular,
they predict that the strength of the Fe line becomes nearly
constant and that the Fe-line equivalent width decreases with
increasing hard X-ray flux for a broad range of power-law indices.
 
In the second explanation \citep{fava2003} it is assumed that the irradiating hard X-ray flux is intrinsically anisotropic. Based on this, \citet{mifago2003} proposed a simple model in which gravitational light
bending creates such an anisotropy. From their study three
regimes of spectral variability are expected: the reflection
component can either be correlated, independent or anti-correlated
with respect to hard X-ray flux, depending on whether the observed hard
X-ray flux is low, medium or high, respectively.  \citet{mifami2004} showed that this light bending model is able to account for variations
in the spectral properties of the black hole candidate XTE J1650--500
as well.  

XTE J1650--500 is a Galactic black-hole X-ray transient
that was in outburst between September 2001 and June 2002. It was the
first transient source to be studied extensively with \xte\  for which simultaneous observations with
higher resolution spectrometers ({\it XMM-Newton} and {\it BeppoSAX})
unambiguously revealed a relativistic Fe emission line. For a description of other aspects of the
outburst we refer to \citet{katoro2003a}, \citet{hoklro2003}, \citet{rohomi2003}, \citet{tokaco2003}, and \citet{tokaka2004}; for details on the relativistically broadened and redshifted Fe emission line we refer to \citet{mifawi2002a} and \citet{mifami2004}.

In order to test the claim by \citet{mifami2004} with more observations we analysed the spectra of 57
{\it RXTE} observations of XTE J1650--500 covering the first 30
days of its 2001/2002 outburst. This is the same period in which
the source showed strong variability and quasi-periodic
oscillations (QPO) in the power density spectra. In the following sections, we detail
our analysis procedure and results.

\section{OBSERVATIONS AND DATA ANALYSIS} 

The 57 pointed {\it RXTE}
observations of XTE J1650--500 used in our study were obtained
between 2001 September 6 and 2001 October 5. 
We used data from the
Proportional Counter Array (PCA) \citep{zhgija1993,jaswgi1996}
and
the High Energy X-ray Timing Experiment (HEXTE) \citep{grblhe1996, roblgr1998}. For the PCA, we only selected data from
proportional counter unit 2 (PCU2). This was the only PCU
active in all observations that, unlike PCU0, did not suffer from
high background. For HEXTE, we used data from both clusters. We extracted 3--25 keV spectra from PCU2 ({\tt Standard 2} mode) and 18--150 keV
spectra from HEXTE (default mode) using tools, calibration software, and background models released with LHEASOFT 5.3.1. The
spectra were background-subtracted and dead time\footnote{Data from all active PCUs were used to estimate the dead time. Even though PCU0 had lost its propane layer, the standard dead time recipe from the {\it RXTE Cook Book} was still approximately valid (Keith Jahoda, private communication), leading to correction on the order of $\sim$1\%.} corrections were
applied. To account for uncertainties in the response matrix, a
systematic error of 0.6\% was added to all PCA spectra. The spectra were fitted using XSPEC v11.2 \citep{ar1996}. 

The
evolution of the spectral and variability properties over the entire
outburst show that XTE J1650--500 underwent two major state transitions \citep{katoro2003a,hoklro2003,rohomi2003}. The period that we selected for our analysis includes  the
transition from the hard state to the soft state, corresponding to groups I/II (2001 Sep.~6 -- 2001 Sep.~20) and III (2001 Sep.~21 -- 2001 Oct.~2) as defined by \citet{hoklro2003} on the basis of variability and X-ray colour properties. During this period, the source was also observed once with {\it
XMM-Newton} (Sep. 13) and three times with {\it BeppoSAX} (Sep. 11, Sep
21, Oct. 3).

\subsection{SPECTRAL FITS}

\begin{figure}
\includegraphics[width=8.5cm]{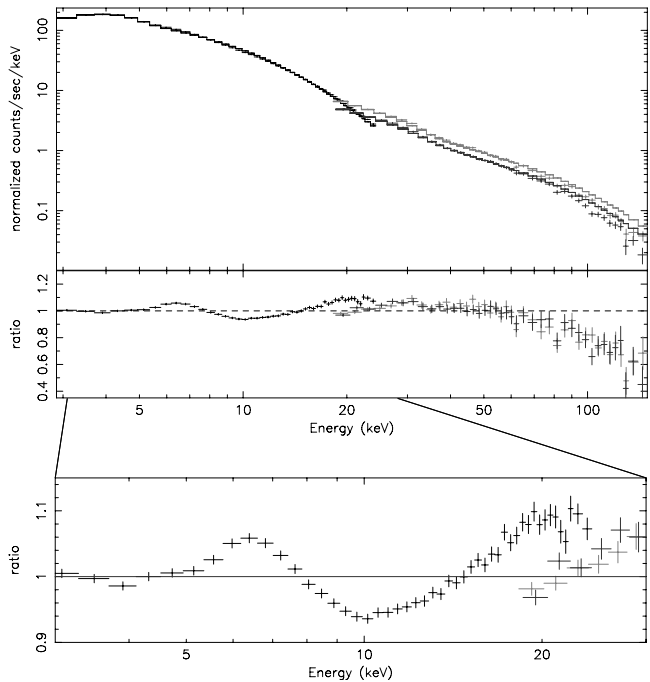}
\caption{An example {\it RXTE} spectrum (PCA in
  black, HEXTE in light and dark grey) of XTE~J1650$-$500, fit with a simple disc
  blackbody plus power-law model.  The data/model ratio traces out the
  curvature predicted when disc reflection is strong.}
\label{fig:fit}
\end{figure}

As we were mainly interested in the behaviour of the Fe line with
respect to general changes in the continuum spectrum, we chose to fit our spectra with the same simple phenomenological continuum components that were used to fit {\it XMM-Newton} and
{\it BeppoSAX} spectra of XTE J1650--500 \citep{mifawi2002a,mifami2004}. These components were a multi-colour discblackbody ({\tt diskbb} in XSPEC), a power-law ({\tt powerlaw}) or cut-off power-law ({\tt cutoff}), and a correction for interstellar absorption ({\tt phabs}). In our case a multiplicative constant was introduced to account for possible mismatches between the PCA and HEXTE spectra. This constant was fixed to 1 for the PCA spectra and had typical values of around 0.88 for the HEXTE spectra. An example fit with this simple model can 
be seen in Figure \ref{fig:fit}.
 In the top panel, we show the 3-160
keV energy spectrum of XTE J1650-500 from 2001 Sep.~13. The ratio between
the data and the model is shown in the middle and bottom panels. The residuals above 50 keV and around 6 keV and 10 keV suggest
the presence of a high-energy cut-off, an Fe~K$\alpha$ emission
line, and an Fe~K absorption edge, respectively.
For a few observations we also tried reflection models ({\tt pexriv} \citep{mazd1995,bama1992} and {\tt cdid} \citep{rofa1993}), which clearly indicated the presence of a reflection component, including the reflection hump. However, fitting with those models proved to be very time consuming and in the case of the {\tt cdid} model no separate line flux could be measured (this is also the case for the {\tt xion} reflection model). In our simplified model, the smeared absorption edge (see below) and the cut-off in the power-law accounted for the reflected continuum.  

In keeping with prior work,
we modelled the Fe~K$\alpha$ emission line using a relativistic Fe
line ({\tt laor}) \citep{la1991} and we also used a smeared edge ({\tt smedge}) due to Fe absorption. For the {\tt laor} component we fixed the outer radius ($ 400
\rm r_{g} $) and the inclination ($45^{\circ}$). The inner disc radius, the index and the normalisation of the emission line were allowed to vary, while the centroid energy was constrained to vary in the 6.4-6.97 keV range. 
For the {\tt smedge} component all the
parameters were left free, except for the index for photo-electric
absorption, which was fixed to its default value of -2.67. The threshold energy was constrained to lie in 7.1--9.3 keV range, the maximum absorption factor was set to be lower than 2.0, and the width to be less than 10.0 keV.

Only in spectra obtained prior
to Sep.~19 a high-energy cut-off in the power-law was required by
the data and for those spectra we used the cut-off power law. The cut-off energy climbed steadily until Sep.~19, and
moved out of our bandpass after Sep.~19. After that day, a simple power law was used. In all our fits the neutral column density $N_{\rm H}$ was fixed to
$5.3 \times 10^{21} $ atoms $ \rm cm^{-2} $ \citep{mifawi2002a,mifami2004}.

Using the above model we were able to get fits with values of $\chi^2$ around 1. However, the results that we obtained for the disc blackbody were not in line with those found from the  {\it BeppoSAX}  and {\it XMM-Newton} spectra that were taken around the same time. Especially in the beginning of group I/II, when the disc blackbody had a relatively small contribution in the {\it RXTE} bandpass \citep[see][]{rohomi2003} and could not be constrained very well, we found high values for the inner disc temperature (kT$\sim$0.8--0.9 keV) and very small values for the inner disc radius (less than 10 km, for an assumed distance of 4 kpc and inclination of 45$^\circ$). 
On Sep.11 and on Sep.13, XTE J1650-500 was observed with {\it BeppoSAX}  and {\it XMM-Newton}, respectively. Both instruments revealed the presence of a disc component in the energy spectrum, and while the values for the {\it BeppoSAX} (0.63$\pm$0.03 keV) and {\it XMM-Newton} (0.322$\pm$0.004 keV) spectra are quite different from each other, they are both considerably lower than the values obtained from our {\it RXTE} spectra. Increasing the lower energy boundary for the {\it XMM-Newton} spectrum to around the lower PCA boundary ($\sim$3 keV) resulted in a fit with a temperature of $\sim$1.2 keV, which suggests that it is the limited sensitivity of the PCA at low energies that gave rise to the anomalous {\tt diskbb} values. For this reason we decided to constrain the inner disc temperature in group I/II to values between 0.3--0.65 keV. In group III our fits gave reasonable disc parameters without these constraints. Using this
model, we obtained values of $\chi_{red}^{2}$ in the range $0.74-1.34$.

\section{RESULTS}

\begin{figure}

\resizebox{!}{1\columnwidth}
{\includegraphics{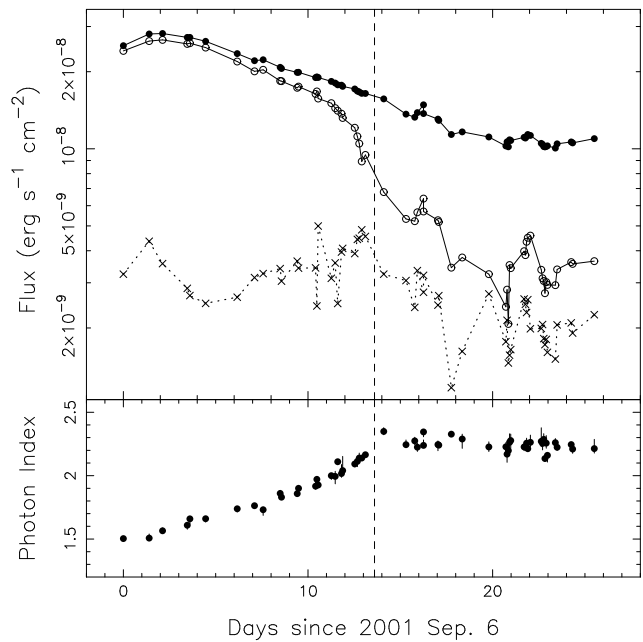}} \caption{Top panel:
evolution of the unabsorbed 2--100 keV total flux (filled circles), the 2--100 
power-law flux (open circles) and the 2--10 keV Fe-line flux (crosses). The Fe-line flux is multiplied by a
factor of 10. Bottom panel: evolution of the power-law photon
index. Errors on the total flux are on the order of a few percent. Errors on the power-law flux are on the order of a few percent for group I/II and around 10 percent for group III. For the errors on on the Fe-line flux refer to Figure \ref{fig:fe_ion}a. The dashed vertical line marks the separation between group I/II and
group III.}\label{fig:fluxes}

\end{figure}

In Figure \ref{fig:fluxes} we present the evolution of the
spectral parameters relevant to our work. The top panel shows the
evolution of the unabsorbed total 2--100 keV flux (filled circles), the
2--100 keV power-law flux (open circles) and the 2--10 keV Fe-line flux (crosses). In the bottom panel the
evolution of the power-law index can be seen. This parameter showed
two distinct trends: between Sep.~6 and Sep.~18, it rose smoothly
from 1.3 to $\sim$2.2--2.3, but from Sep.~19 onwards it remained almost constant around this level. This change in behavior coincided with a change in the variability properties \citep{hoklro2003,rohomi2003}.

At the start of our dataset the total flux was dominated by the power-law component:
at its peak ( $2.8 \times
10^{-8}$ \ergpscmps) the contribution of the power-law flux was around
94\% of the total flux. In subsequent observations the total flux decayed and
the fractional contribution of the power-law component gradually decreased. On Sep.~18, the power-law flux started to decay more rapidly. At that point it had already had dropped to 36\% of its peak value. The rapid decay continued for a few more days, until the power-law flux started to vary erratically between $\sim$2$\times10^{-9}$ and $\sim$5$\times10^{-9}$ \ergpscmps  with a fractional contribution of $\sim$20--40\%. The Fe-line flux always contributed around 1--2\%
of the total flux; the remaining flux was due to the disc
component, which dominated after Sep.~19. From Figure \ref{fig:fluxes}  one can see that before
Sep. 19 the flux evolution of the Fe line did not follow that of the
power-law, with a negative correlation coefficient of --0.54, whereas after that day there was evidence of a positive
correlation (coefficient of 0.80).

Since XTE J1650--500 behaved rather differently in the time
intervals before Sep. 18 and after Sep. 19 we will, in the following, discuss the observations in terms of the two corresponding groups that were introduced by \citet{hoklro2003}, i.e. groups I/II and III, respectively.

\subsection{Relationship between the power-law and Fe-line fluxes}

\begin{figure}
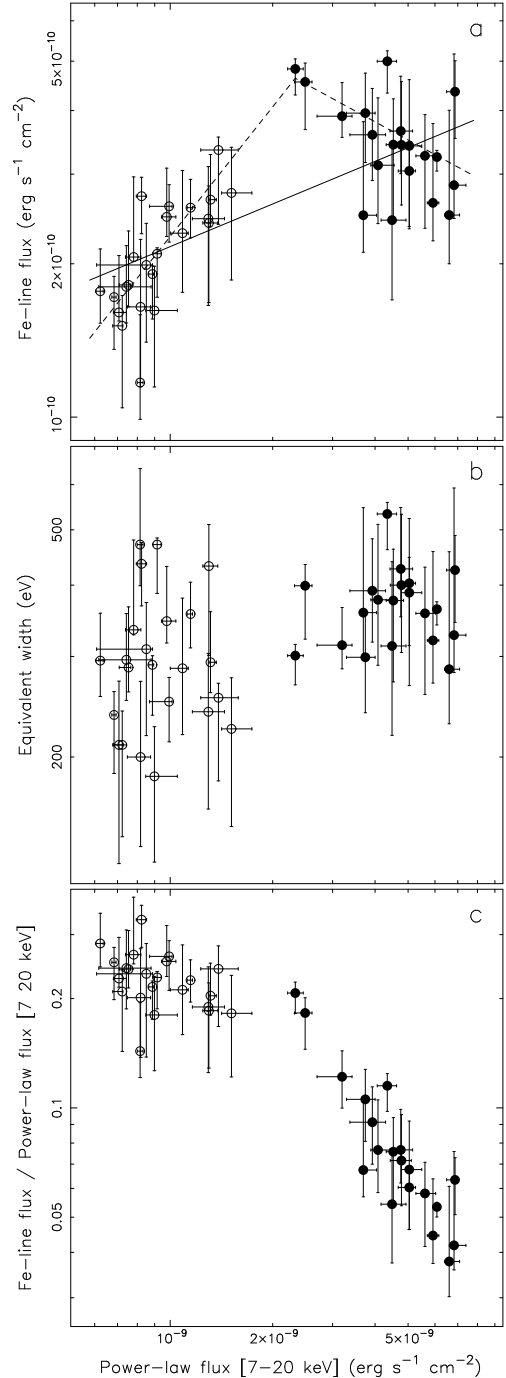


\centerline{\includegraphics[width=6.5cm]{f3a.eps}}
\centerline{\includegraphics[width=6.5cm]{f3b.eps}}
\centerline{\includegraphics[width=6.5cm]{f3c.eps}}
\caption{
Fe-line properties as a function of the
power-law flux in the 7--20 keV energy band. Line-flux is shown in (a), the equivalent width in (b) and the ratio of line and power-law fluxes in (c). The errors on the line parameters are propagated from the $1\sigma$ error
on the Fe-line normalisation. Only the points with $\rm N_{Fe}/
\Delta N_{Fe} \ge 3.0$ are shown.  Filled circles are used for
group I/II,  open circles for group III. The solid and dashed lines in panel (a) show the best fits to the data with a power-law and broken power-law, respectively.}\label{fig:fe_ion}
\end{figure}

Within the framework of reflection models, the power-law  and Fe-line
flux variations trace the spectral evolution of the irradiating flux and the
reflection component, respectively. To explore the
relationship between these two components in XTE J1650--500 in more detail, we
obtained the power-law flux in the 7-20 keV energy band (hereafter $F_{7-20}$) from our
fits. This flux represents the hard X-ray photons
which can cause inner-shell fluorescence. The fraction of this flux with respect to the
total 2--100 keV power-law flux was $\sim$0.24--0.28. In
Figure \ref{fig:fe_ion} we show the Fe-line flux (\ref{fig:fe_ion}a) and the Fe-line equivalent
width (\ref{fig:fe_ion}b) as a function of $F_{7-20}$, using
filled circles for group I/II and open circles for group III.
The fractional error on the Fe-line flux was set equal to the
$1\sigma $ fractional error on the Fe-line normalisation, $\rm N_{Fe}$. We only show data points for which $\rm N_{Fe}/ \Delta
N_{Fe} \ge 3.0$, with $\rm \Delta N_{Fe}$ the negative 1$\sigma$ error on the
normalisation.

As already suggested by Figure \ref{fig:fluxes}, we found that the irradiating flux and the Fe-line flux did not correlate in a simple linear way. A linear fit to the data in Figure \ref{fig:fe_ion}a, although presenting a clear improvement over a fit with a constant ($\chi^2/d.o.f.$=185/42), only yielded a $\chi^2/d.o.f.$ of 100/41. A power-law fit gave a slightly better $\chi^2/d.o.f.$ of 82/41, but both this fit and the linear one had problems to fit the apparent change in the Fe-line/power-law flux behaviour between the two groups. Qualitatively, in group I/II, the line flux showed
little variations (or maybe an anti-correlation - correlation coefficient is --0.54) over a large range of $F_{7-20}$, while there
are indications for a positive correlation in group III (correlation coefficient is 0.8). To allow for different behaviour in the two groups we used a broken power-law, which resulted in a $\chi^2/d.o.f.$ of 33/39, with a break at a flux of 2.3$\pm$0.1$\times10^{-9}$ \ergpscmps and indices of 0.8$\pm$0.1 and $-0.4\pm$0.1 below and above the break, respectively. An F-test shows that this fit is better than a single power-law fit with a probability of 99.996\%. A broken line gives an equally good fit, while a second order polynomial results in $\chi^2/d.o.f.$=52/40. Both the broken line fit and broken power-law fit suggest that in group I/II the line flux may be anti-correlated with the irradiating flux. The best fits with a power law and broken power-law are shown in Figure \ref{fig:fe_ion}a.

In Figure \ref{fig:fe_ion}b we show the line equivalent width as a function of $F_{7-20}$. The equivalent width had values between $\sim$200 eV and $\sim$500 eV, with indications for only a moderate increase with $F_{7-20}$. Fits with a constant to group I/II, group III, and all data combined resulted in values of  357$\pm$11 eV,  283$\pm$11 eV, and 320$\pm$8, respectively.

In some of the works that are relevant to the discussion in the next section the equivalent width is calculated from simulated spectra in which the continuum is dominated by a power-law component, either because a disc is not part of the model or the disc does not contribute much in X-rays. In our spectra, however, a prominent (and evolving) disc component is present, especially at the end of group I/II and in group III. We found that changes in the continuum flux around 6--7 keV due to a decrease in the power-law component were (partly) compensated by a simultaneous increase in the disc flux. Hence, the behaviour of our measured equivalent width cannot be directly compared to that from simulations in which there is no other component to compensate changes in the power-law flux. We therefore calculated the flux ratio of the Fe-line and the power-law components, as a measure of the Fe-line equivalent width with respect to the power-law component. The results are shown in Figure \ref{fig:fe_ion}c and revealed strong decrease in the relative importance of the line at higher values of the power-law flux.

The errors on the Fe-line energy were too large to study its evolution during our observations. The emissivity index of the Laor component varied between 4 and 6 but was in most cases consistent with the value of 5.4 that was found by \citet{mifawi2002a}.

\section{DISCUSSION}

We have studied the variability of the
Fe~K$\alpha$ emission line and power-law flux components in {\it RXTE}
spectra of the Galactic black-hole candidate XTE~J1650$-$500.  This is
only the second time that a large number of \xte\ observations of a transient outburst has been analyzed using a
relativistic model for the Fe~K$\alpha$ emission line and a continuum
that approximates a full reflection model (\citet{pamimc2004} analyzed
{\it RXTE} observations of the 2002 outburst of 4U~1543$-$475 with a
relativistic line model).  

These studies of 4U 1543$-$475 and XTE~J1650$-$500 represent
particularly important tests of disc reflection models.  Simple disc
reflection models \citep[e.g.][]{gefa1991} predict a positive
linear correlation between the flux of the power-law component
(assumed to irradiate the disc) and the flux of the line component.
Similar to the case of 4U~1543$-$475, we do not observe a simple,
positive linear correlation between line flux and power-law flux.
Unlike the case of 4U~1543$-$475, however, we observe a clear and
continuous trend: at low power-law flux levels, the line and power-law
fluxes are positively correlated; at high power-law flux levels the
line flux is possibly inversely related.

New disc reflection models which incorporate the effects of disc
ionization \citep[e.g.][]{baro2002} and realistic disc
atmospheres \citep[e.g.][]{naka2001} do not necessarily
predict simple positive correlations between line flux and power-law
flux.  The expected line flux depends not only on the ionisation state
of the disc, but on the hardness of the flux which irradiates the disc
as well. \citet{baro2002} predict that line flux becomes
nearly constant with increasing power-law flux, and that line
equivalent width decreases with increasing power-law flux for a broad
range of power-law indices.  These predictions are in broad agreement
with our results (assuming that the flux ratio plotted in Fig.~\ref{fig:fe_ion}c provides a better means of comparison than the equivalent width plotted in Fig.~\ref{fig:fe_ion}b, which is diluted by the disc component).

\citet{mifa2004} have developed a reflection model which
incorporates the effects of light bending close to a black hole (but does not take into account the effects of disc ionisation).  This
model predicts that the flux of Fe~K$\alpha$ emission line is positively
correlated with the power-law flux at low values of the power-law fluxes, constant or
insensitive to power-law flux at moderate power-law fluxes, and
anti-correlated with power-law flux at high power-law fluxes.  The
anti-correlation is an important distinction between this model and
reflection models which focus on disc ionisation effects.  This model
also predicts that line equivalent width and power-law flux should be
anti-correlated, which seems to be confirmed by Figure \ref{fig:fe_ion}c.  \citet{mifami2004} have noted that
the Fe~K$\alpha$ emission line flux versus power-law flux trend seen
in XTE J1650$-$500 is in broad agreement with the light bending model;
however, that work was based on only three observations obtained with
{\it BeppoSAX}.  Our analysis of 57 pointed {\it RXTE} observations
better reveals that the line versus power-law flux and line equivalent
width versus power-law flux trends seen in XTE~J1650$-$500 are in
broad agreement also with the predictions of the light bending model.  

Thus, our results are broadly consisent with predictions made by both
the ionised disc reflection models \citep[e.g.][]{baro2002}, and
a model which focuses on light bending effects \citep{mifa2004}.  We find evidence for an anti-correlation between line
flux and power-law flux at high power-law flux values.  This may lend
support to the light bending model.  Moreover, fits to spectra of
XTE~J1650$-$500 obtained with {\it XMM-Newton} and {\it BeppoSAX} --
which have the energy resolution required to reliably measure the Fe
charge state -- show that the line is broadly consistent with He-like Fe~XXV over a period of more than twenty days covering groups I/II and III.
This may indicate that the changing disc ionisation
was not responsible for the line variability observed in
XTE~J1650$-$500.  Taken together, these facts lend some additional
support to the light bending model, but are not decisive.  We briefly
discuss the light bending model for
XTE~J1650$-$500 below, because the model is particularly interesting
and offers some novel physical constraints.  This discussion should
not be taken as a judgement that the data clearly rule in favor of the
light bending model.

It is reasonable to expect relativistic
effects to be of importance in XTE J1650--500, with \citet{mifawi2002a} and \citet{mifami2004} having shown that the Fe line in this source could
originate from as close as 2$r_g$ from the black hole. In the
framework of the light-bending model the variability of the hard and reflected
components observed in spectra from the early phase of the 2001/2002
outburst of XTE J1650--500 can be explained by a variable height of the hard
X-ray source above the accretion disc: in the beginning, the hard
X-ray source is located at a medium distance (group I/II), then it
gradually approaches to the disc reaching very short distances after
Sep.19 (group III).  Using the predictions of \citet{mifa2004} and assuming that XTE J1650--500 has an inclination of $45^{\circ}$, the height of the hard X-ray source in group I/II is in
the range of $2-4 r_g < h_s < 7-13 r_{g}$ and reaches values lower
than $\sim 2-4 r_{g}$ in group III.

The explanation of such height variations obviously depends on the model that one assumes for the spectrally hard component in black-hole X-ray binaries. A detailed discussion of all proposed models is beyond the scope of this paper, but we wish to briefly mention the jet model by \citet{mafafe2001} \citep[see also][]{manoco2003b}, because radio observations of XTE J1650--500 show evidence for the presence of a strong compact jet \citep{cofeto2004}. If the
hard flux in accreting black holes is produced in part by the base of
a jet, then the height of the jet base may act to set a size scale
for the corona and reasonably be approximated by a central
"lamp-post" source like the idealised hard source assumed in the
light-bending model. The radio observations of XTE J1650--500 suggest a moderate evolution in the jet properties from I/II to group III, which may then be related to changes in the scale height of the base of the jet.

The light bending model also predicts
that differing degrees of light bending can cause a hard component
with a {\it constant} flux to give the appearance of a component
which varies in flux by a factor of $\sim$20 (depending on
inclination), and indeed the hard X-ray flux variations we observe
in XTE~J1650$-$500 only span a factor of $\sim$10.
It is undoubtedly the case that variations in the mass accretion rate
must play a central role in transient outbursts, and our results in no
way serve to contradict that premise.  Merely, they suggest
that moderate variations in the strength of the hard X-ray emission
during state transitions may be
partially due to light bending effects.  

It is worth noting the change in the Fe-line behaviour occurs during the transition from the hard to the soft state. In particular, the point at which the Fe line variations
change regime (Sep. 19) coincides with the point at which the
power-law photon index becomes relatively constant, and
the point at which 250~Hz QPOs (presumably from very close to the black hole) are first observed (Homan et al.\
2003). Detailed timing analysis of the {\it RXTE} dataset \citep{hoklro2003,rohomi2003} reveals
that the frequency of QPOs below 10~Hz climbs steadily until Sep.
19, and then reaches a more-or-less steady value after Sep. 19.  
Thus, it might be
possible that light-bending and/or disc-ionization effects are not only partially contributing to the
Fe-line flux variations, but also timing properties.  

Clearly, despite its limited spectral resolution, {\it RXTE} can make
important contributions to understanding Galactic black-hole spectra,
and can reveal possible relativistic effects which are manifested
spectrally. The scheduling flexibility of {\it RXTE} means that it is
the {\it only} X-ray observatory, present or planned, that can
execute the numerous observations required to reveal evidence of
effects like gravitational light bending.  Analysis of archival data and future dedicated observing efforts with {\it RXTE} can be used to search for improved
evidence of light bending and other relativistic effects in Galactic
black holes.

\section*{Acknowledgements} The authors would like to thank the referee for constructive comments on an earlier version of the manuscript. S.\ Rossi thanks G.\ Ghisellini, F.\
Tavecchio and S.\ Campana for useful discussions.   J.\ Miller and J.\
Homan thank R.\ Remillard, M.\ Nowak, and G.\ Miniutti for useful
discussions.  S.\ Rossi gratefully acknowledges financial support from ASI and MIT.  J.\ Miller gratefully acknowledges support from the U.
S. National Science Foundation through its Astronomy and Astrophysics
Postdoctoral Fellowship program. J.\ Homan gratefully acknowledges
support from NASA.


\begin{thebibliography}{}

\bibitem[\protect\citeauthoryear{{Arnaud}}{{Arnaud}}{1996}]{ar1996}
{Arnaud} K.~A.,  1996, in ASP Conf. Ser. 101: Astronomical Data Analysis
  Software and Systems V Vol.~5, Xspec: The first ten years.
p.~17

\bibitem[\protect\citeauthoryear{{Ballantyne} \& {Ross}}{{Ballantyne} \&
  {Ross}}{2002}]{baro2002}
{Ballantyne} D.~R.,  {Ross} R.~R.,  2002, MNRAS, 332, 777

\bibitem[\protect\citeauthoryear{{Ballantyne}, {Ross} \& {Fabian}}{{Ballantyne}
  et~al.}{2001}]{barofa2001}
{Ballantyne} D.~R.,  {Ross} R.~R.,    {Fabian} A.~C.,  2001, MNRAS, 327, 10

\bibitem[\protect\citeauthoryear{{Ballantyne}, {Vaughan} \&
  {Fabian}}{{Ballantyne} et~al.}{2003}]{bavafa2003}
{Ballantyne} D.~R.,  {Vaughan} S.,    {Fabian} A.~C.,  2003, MNRAS, 342, 239

\bibitem[\protect\citeauthoryear{{Balucinska-Church} \&
  {McCammon}}{{Balucinska-Church} \& {McCammon}}{1992}]{bama1992}
{Balucinska-Church} M.,  {McCammon} D.,  1992, ApJ, 400, 699

\bibitem[\protect\citeauthoryear{{Corbel}, {Fender}, {Tomsick}, {Tzioumis} \&
  {Tingay}}{{Corbel} et~al.}{2004}]{cofeto2004}
{Corbel} S.,  {Fender} R.~P.,  {Tomsick} J.~A.,  {Tzioumis} A.~K.,    {Tingay}
  S.,  2004, ApJ, 617, 1272

\bibitem[\protect\citeauthoryear{{Fabian} \& {Vaughan}}{{Fabian} \&
  {Vaughan}}{2003}]{fava2003}
{Fabian} A.~C.,  {Vaughan} S.,  2003, MNRAS, 340, L28

\bibitem[\protect\citeauthoryear{{Fabian}, {Vaughan}, {Nandra}, {Iwasawa},
  {Ballantyne}, {Lee}, {De Rosa}, {Turner} \& {Young}}{{Fabian}
  et~al.}{2002}]{favana2002}
{Fabian} A.~C.,  {Vaughan} S.,  {Nandra} K.,  {Iwasawa} K.,  {Ballantyne}
  D.~R.,  {Lee} J.~C.,  {De Rosa} A.,  {Turner} A.,    {Young} A.~J.,  2002,
  MNRAS, 335, L1

\bibitem[\protect\citeauthoryear{{George} \& {Fabian}}{{George} \&
  {Fabian}}{1991}]{gefa1991}
{George} I.~M.,  {Fabian} A.~C.,  1991, MNRAS, 249, 352

\bibitem[\protect\citeauthoryear{{Gruber}, {Blanco}, {Heindl}, {Pelling},
  {Rothschild} \& {Hink}}{{Gruber} et~al.}{1996}]{grblhe1996}
{Gruber} D.~E.,  {Blanco} P.~R.,  {Heindl} W.~A.,  {Pelling} M.~R.,
  {Rothschild} R.~E.,    {Hink} P.~L.,  1996, A\&A Supp., 120, C641

\bibitem[\protect\citeauthoryear{{Homan}, {Klein-Wolt}, {Rossi}, {Miller},
  {Wijnands}, {Belloni}, {van der Klis} \& {Lewin}}{{Homan}
  et~al.}{2003}]{hoklro2003}
{Homan} J.,  {Klein-Wolt} M.,  {Rossi} S.,  {Miller} J.~M.,  {Wijnands} R.,
  {Belloni} T.,  {van der Klis} M.,    {Lewin} W.~H.~G.,  2003, ApJ, 586, 1262

\bibitem[\protect\citeauthoryear{{Jahoda}, {Swank}, {Giles}, {Stark},
  {Strohmayer}, {Zhang} \& {Morgan}}{{Jahoda} et~al.}{1996}]{jaswgi1996}
{Jahoda} K.,  {Swank} J.~H.,  {Giles} A.~B.,  {Stark} M.~J.,  {Strohmayer} T.,
  {Zhang} W.,    {Morgan} E.~H.,  1996, Proc. SPIE, 2808, 59

\bibitem[\protect\citeauthoryear{{Kalemci}, {Tomsick}, {Rothschild},
  {Pottschmidt}, {Corbel}, {Wijnands}, {Miller} \& {Kaaret}}{{Kalemci}
  et~al.}{2003}]{katoro2003a}
{Kalemci} E.,  {Tomsick} J.~A.,  {Rothschild} R.~E.,  {Pottschmidt} K.,
  {Corbel} S.,  {Wijnands} R.,  {Miller} J.~M.,    {Kaaret} P.,  2003, ApJ,
  586, 419

\bibitem[\protect\citeauthoryear{{Laor}}{{Laor}}{1991}]{la1991}
{Laor} A.,  1991, ApJ, 376, 90

\bibitem[\protect\citeauthoryear{{Lee}, {Fabian}, {Reynolds}, {Brandt} \&
  {Iwasawa}}{{Lee} et~al.}{2000}]{lefare2000}
{Lee} J.~C.,  {Fabian} A.~C.,  {Reynolds} C.~S.,  {Brandt} W.~N.,    {Iwasawa}
  K.,  2000, MNRAS, 318, 857

\bibitem[\protect\citeauthoryear{{Magdziarz} \& {Zdziarski}}{{Magdziarz} \&
  {Zdziarski}}{1995}]{mazd1995}
{Magdziarz} P.,  {Zdziarski} A.~A.,  1995, MNRAS, 273, 837

\bibitem[\protect\citeauthoryear{{Markoff}, {Falcke} \& {Fender}}{{Markoff}
  et~al.}{2001}]{mafafe2001}
{Markoff} S.,  {Falcke} H.,    {Fender} R.,  2001, A\&A, 372, L25

\bibitem[\protect\citeauthoryear{{Markoff}, {Nowak}, {Corbel}, {Fender} \&
  {Falcke}}{{Markoff} et~al.}{2003}]{manoco2003b}
{Markoff} S.,  {Nowak} M.,  {Corbel} S.,  {Fender} R.,    {Falcke} H.,  2003,
  New Astronomy Review, 47, 491

\bibitem[\protect\citeauthoryear{{Martocchia}, {Matt} \& {Karas}}{{Martocchia}
  et~al.}{2002}]{mamaka2002}
{Martocchia} A.,  {Matt} G.,    {Karas} V.,  2002, A\&A, 383, L23

\bibitem[\protect\citeauthoryear{{Miller}, {Fabian}, {in't Zand}, {Reynolds},
  {Wijnands}, {Nowak} \& {Lewin}}{{Miller} et~al.}{2002b}]{mifain2002}
{Miller} J.~M.,  {Fabian} A.~C.,  {in't Zand} J.~J.~M.,  {Reynolds} C.~S.,
  {Wijnands} R.,  {Nowak} M.~A.,    {Lewin} W.~H.~G.,  2002, ApJ, 577, L15

\bibitem[\protect\citeauthoryear{{Miller}, {Fabian}, {Wijnands}, {Remillard},
  {Wojdowski}, {Schulz}, {Di Matteo}, {Marshall}, {Canizares}, {Pooley} \&
  {Lewin}}{{Miller} et~al.}{2002c}]{mifawi2002b}
{Miller} J.~M.,  {Fabian} A.~C.,  {Wijnands} R.,  {Remillard} R.~A.,
  {Wojdowski} P.,  {Schulz} N.~S.,  {Di Matteo} T.,  {Marshall} H.~L.,
  {Canizares} C.~R.,  {Pooley} D.,    {Lewin} W.~H.~G.,  2002, ApJ, 578, 348

\bibitem[\protect\citeauthoryear{{Miller}, {Fabian}, {Wijnands}, {Reynolds},
  {Ehle}, {Freyberg}, {van der Klis}, {Lewin}, {S\'anchez-Fern\'andez} \&
  {Castro-Tirado}}{{Miller} et~al.}{2002a}]{mifawi2002a}
{Miller} J.~M.,  {Fabian} A.~C.,  {Wijnands} R.,  {Reynolds} C.~S.,  {Ehle} M.,
   {Freyberg} M.~J.,  {van der Klis} M.,  {Lewin} W.~H.~G.,
  {S\'anchez-Fern\'andez} C.,    {Castro-Tirado} A.~J.,  2002, ApJ, 570, L69

\bibitem[\protect\citeauthoryear{{Miniutti} \& {Fabian}}{{Miniutti} \&
  {Fabian}}{2004}]{mifa2004}
{Miniutti} G.,  {Fabian} A.~C.,  2004, MNRAS, 349, 1435

\bibitem[\protect\citeauthoryear{{Miniutti}, {Fabian}, {Goyder} \&
  {Lasenby}}{{Miniutti} et~al.}{2003}]{mifago2003}
{Miniutti} G.,  {Fabian} A.~C.,  {Goyder} R.,    {Lasenby} A.~N.,  2003,
  MNRAS, 344, L22

\bibitem[\protect\citeauthoryear{{Miniutti}, {Fabian} \& {Miller}}{{Miniutti}
  et~al.}{2004}]{mifami2004}
{Miniutti} G.,  {Fabian} A.~C.,    {Miller} J.~M.,  2004, MNRAS, 351, 466

\bibitem[\protect\citeauthoryear{{Nayakshin} \& {Kallman}}{{Nayakshin} \&
  {Kallman}}{2001}]{naka2001}
{Nayakshin} S.,  {Kallman} T.~R.,  2001, ApJ, 546, 406

\bibitem[\protect\citeauthoryear{{Nayakshin}, {Kazanas} \&
  {Kallman}}{{Nayakshin} et~al.}{2000}]{nakaka2000}
{Nayakshin} S.,  {Kazanas} D.,    {Kallman} T.~R.,  2000, ApJ, 537, 833

\bibitem[\protect\citeauthoryear{{Park}, {Miller}, {McClintock}, {Remillard},
  {Orosz}, {Shrader}, {Hunstead}, {Campbell-Wilson}, {Ishwara-Chandra}, {Rao}
  \& {Rupen}}{{Park} et~al.}{2004}]{pamimc2004}
{Park} S.~Q.,  {Miller} J.~M.,  {McClintock} J.~E.,  {Remillard} R.~A.,
  {Orosz} J.~A.,  {Shrader} C.~R.,  {Hunstead} R.~W.,  {Campbell-Wilson} D.,
  {Ishwara-Chandra} C.~H.,  {Rao} A.~P.,    {Rupen} M.~P.,  2004, ApJ, 610,
  378

\bibitem[\protect\citeauthoryear{{Reynolds} \& {Nowak}}{{Reynolds} \&
  {Nowak}}{2003}]{reno2003}
{Reynolds} C.~S.,  {Nowak} M.~A.,  2003, Phys. Rep., 377, 389

\bibitem[\protect\citeauthoryear{{Ross} \& {Fabian}}{{Ross} \&
  {Fabian}}{1993}]{rofa1993}
{Ross} R.~R.,  {Fabian} A.~C.,  1993, MNRAS, 261, 74

\bibitem[\protect\citeauthoryear{{Ross}, {Fabian} \& {Young}}{{Ross}
  et~al.}{1999}]{rofayo1999}
{Ross} R.~R.,  {Fabian} A.~C.,    {Young} A.~J.,  1999, MNRAS, 306, 461

\bibitem[\protect\citeauthoryear{{Rossi}, {Homan}, {Miller} \&
  {Belloni}}{{Rossi} et~al.}{2003}]{rohomi2003}
{Rossi} S.,  {Homan} J.,  {Miller} J.,    {Belloni} T.,  2003, astro-ph/0309129

\bibitem[\protect\citeauthoryear{{Rothschild}, {Blanco}, {Gruber}, {Heindl},
  {MacDonald}, {Marsden}, {Pelling}, {Wayne} \& {Hink}}{{Rothschild}
  et~al.}{1998}]{roblgr1998}
{Rothschild} R.~E.,  {Blanco} P.~R.,  {Gruber} D.~E.,  {Heindl} W.~A.,
  {MacDonald} D.~R.,  {Marsden} D.~C.,  {Pelling} M.~R.,  {Wayne} L.~R.,
  {Hink} P.~L.,  1998, ApJ, 496, 538

\bibitem[\protect\citeauthoryear{{Shih}, {Iwasawa} \& {Fabian}}{{Shih}
  et~al.}{2002}]{shiwfa2002}
{Shih} D.~C.,  {Iwasawa} K.,    {Fabian} A.~C.,  2002, MNRAS, 333, 687

\bibitem[\protect\citeauthoryear{{Tanaka}, {Nandra}, {Fabian}, {Inoue},
  {Otani}, {Dotani}, {Hayashida}, {Iwasawa}, {Kii}, {Kunieda}, {Makino} \&
  {Matsuoka}}{{Tanaka} et~al.}{1995}]{tanafa1995}
{Tanaka} Y.,  {Nandra} K.,  {Fabian} A.~C.,  {Inoue} H.,  {Otani} C.,  {Dotani}
  T.,  {Hayashida} K.,  {Iwasawa} K.,  {Kii} T.,  {Kunieda} H.,  {Makino} F.,
   {Matsuoka} M.,  1995, Nature, 375, 659

\bibitem[\protect\citeauthoryear{{Tomsick}, {Kalemci}, {Corbel} \&
  {Kaaret}}{{Tomsick} et~al.}{2003}]{tokaco2003}
{Tomsick} J.~A.,  {Kalemci} E.,  {Corbel} S.,    {Kaaret} P.,  2003, ApJ, 592,
  1100

\bibitem[\protect\citeauthoryear{{Tomsick}, {Kalemci} \& {Kaaret}}{{Tomsick}
  et~al.}{2004}]{tokaka2004}
{Tomsick} J.~A.,  {Kalemci} E.,    {Kaaret} P.,  2004, ApJ, 601, 439

\bibitem[\protect\citeauthoryear{{Zhang}, {Giles}, {Jahoda}, {Soong}, {Swank}
  \& {Morgan}}{{Zhang} et~al.}{1993}]{zhgija1993}
{Zhang} W.,  {Giles} A.~B.,  {Jahoda} K.,  {Soong} Y.,  {Swank} J.~H.,
  {Morgan} E.~H.,  1993, Proc. SPIE, 2006, 324

\end{thebibliography}

\bsp
\label{lastpage}
\end{document}